\documentclass[preprint,preprintnumbers,nofootinbib,tightenlines,12pt,superscriptaddress]{revtex4}

\RequirePackage[12tabu, orthodox]{nag}
\usepackage[all, warning]{onlyamsmath}

\usepackage{amsmath,amssymb,setspace,amsfonts,latexsym,braket}
\usepackage[colorlinks]{hyperref}
\usepackage{graphicx}
\usepackage{color}

\newcommand\bea{\begin{eqnarray}}
\newcommand\eea{\end{eqnarray}}

\def\vrel{v_{\rm rel}}
\def\semi{\sigma_{\rm semi} \vrel}
\def\invsemi{\sigma_{\rm inv} \vrel}

\def\self{\sigma_{\rm self} \vrel}

\newcommand\unit[1]{\,{\rm #1}}
\newcommand\eV{\unit{eV}}
\newcommand\keV{\unit{keV}}
\newcommand\MeV{\unit{MeV}}
\newcommand\GeV{\unit{GeV}}

\begin{document}

\hfill CTPU-PTC-18-11

\title{\Large Self-heating of Strongly Interacting Massive Particles}
\author{Ayuki Kamada} 
\email{akamada@ibs.re.kr}
\affiliation{Center for Theoretical Physics of the Universe, Institute for Basic Science (IBS), Daejeon 34126, Korea}
\author{Hee Jung Kim} 
\email{hyzer333@kaist.ac.kr}
\affiliation{Department of Physics, KAIST, Daejeon 34141, Korea}
\author{Hyungjin Kim}
\email{hyungjin.kim@weizmann.ac.il}
\affiliation{Department of Particle Physics and Astrophysics, Weizmann Institute of Science, Rehovot 7610001, Israel\bigskip}

\begin{abstract}
It was recently pointed out that semi-annihilating dark matter (DM) may experience a novel temperature evolution dubbed as {\it self-heating}.
Exothermic semi-annihilation converts the DM mass to the kinetic energy.
This yields a unique DM temperature evolution, $T_{\chi} \propto 1 / a$, in contrast to $ T_{\chi} \propto 1 / a^{2}$ for free-streaming non-relativistic particles.
Self-heating continues as long as self-scattering sufficiently redistributes the energy of DM particles.
In this paper, we study the evolution of cosmological perturbations in self-heating DM.
We find that sub-GeV self-heating DM leaves a cutoff on the subgalactic scale of the matter power spectrum when the self-scattering cross section is $\sigma_{\rm self} / m_{\chi} \sim {\cal O} (1) \unit{cm^{2} / g}$.
Then we present a particle physics realization of the self-heating DM scenario.
The model is based on recently proposed strongly interacting massive particles with pion-like particles in a QCD-like sector.
Pion-like particles semi-annihilate into an axion-like particle, which is thermalized with dark radiation.
The dark radiation temperature is smaller than the standard model temperature, evading the constraint from the effective number of neutrino degrees of freedom.
It is easily realized when the dark sector is populated from the standard model sector through a small coupling.
\end{abstract}

\maketitle

\newpage

\section{Introduction}
Accumulated observational data begin to test our understanding of how dark matter (DM) is distributed over the Universe and how the DM distribution evolves in time~\cite{Buckley:2017ijx}.
In regard to the structure formation of the Universe, DM is often described as a perfect fluid with zero temperature, which is referred to as cold dark matter (CDM).
The CDM paradigm is successful in reproducing the large-scale structure of the Universe observed through cosmic microwave background (CMB) anisotropies~\cite{Ade:2015xua} and galaxy clustering~\cite{Alam:2016hwk}.

Contrary to its success on large scales, CDM predictions appear to be incompatible with the observations on smaller scales~\cite{Bullock:2017xww}.
Although baryonic physics may play an important role~\cite{Sawala:2015cdf, Dutton:2015nvy, Wetzel:2016wro}, these small scale issues may be hinting to alternative DM models.
One example of the CDM failure is the {\it missing satellite problem}, indicating that CDM overpredicts the number of dwarf-size subhalos in a Milky Way-size halo when compared to that of the observed satellite galaxies~\cite{Moore:1999nt, Kravtsov:2009gi}.
Warm dark matter (WDM) is an interesting possibility in this respect. 
Gravitational clustering of WDM particles is interrupted on a subgalactic scale because of a sizable thermal velocity of $v / c \sim 10^{- 3} \text{--} 10^{-4}$ at the matter-radiation equality.
This suppresses dwarf-size halo formation~\cite{Bode:2000gq, Lovell:2011rd}. 
A sterile neutrino with a keV mass is a good benchmark model of WDM, where its phenomenology is described by the mass and the mixing angle with an active neutrino in the simplest setup~\cite{Adhikari:2016bei}.

Another example is the {\it core-cusp problem}: Some dwarf-size halos have a cuspy profile as predicted by CDM, while others have a cored profile~\cite{Moore:1999gc, deBlok:2009sp, Oman:2015xda}.
Self-interacting dark matter (SIDM) is an interesting solution.
The self-scattering cross section of $\sigma_{\rm self} / m_{\chi} \sim 1 \unit{cm^{2} / g}$ with the DM mass $m_{\chi}$ leads to iso-thermalization of DM particles, whose distribution is characterized by a kpc core~\cite{Spergel:1999mh, Rocha:2012jg}.
SIDM reproduces the observed diversity of the rotation velocity among similar-size halos by adjusting its distribution sensitively to the baryon distribution~\cite{Kaplinghat:2015aga, Kamada:2016euw}.

A new possibility, which is called self-heating DM, has been proposed recently~\cite{Kamada:2017gfc}.
A characteristic feature of this scenario is that the strength of self-scattering is related to the thermal velocity of DM and, hence, potentially solves several small scale issues simultaneously.
The original proposal was based on semi-annihilating DM~\cite{Hambye:2008bq,DEramo:2010keq, Belanger:2012vp}, $\chi \chi \to \chi \phi$ with a light particle $\phi$ in the thermal bath.
A key observation is that semi-annihilation converts the mass of DM into the kinetic energy, which leads to the novel DM temperature evolution, $T_{\chi} \propto 1 / a$ after the freeze-out, instead of $T_{\chi} \propto 1 / a^{2}$ for free-streaming non-relativistic particles.
Self-heating continues as long as self-scattering occurs rapidly.
DM self-scattering is an essential ingredient of the scenario because it redistributes the large kinetic energy of the boosted DM particles through the semi-annihilation over the whole DM particles.
Stronger self-scattering elongates the duration of self-heating, resulting in a larger thermal motion of DM particles.

In this paper, we investigate the impact of DM self-heating on the  matter distribution of the Universe, and propose a viable particle physics realization of self-heating DM. 
To study the structure formation, we derive the evolution equation of the cosmological perturbations in the self-heating DM scenario.
We show that semi-annihilation not only changes the DM temperature evolution but also affects the entropic perturbation of DM.
We follow the evolution of the primordial perturbations numerically and show that self-heating DM can leave a subgalactic-scale cutoff in the linear matter power spectrum.
We extend the recently proposed strongly interacting massive particle (SIMP) model~\cite{Carlson:1992fn, Hochberg:2014dra} to realize self-heating DM.
Pion-like particles in a QCD-like sector semi-annihilate into an axion-like particle (ALP), which is thermalized with dark radiation.
We identify a parameter region that is compatible with observational constraints.

This paper is organized as follows.
In Sec.~\ref{sec:self-heat_DM}, we describe the thermal history of the self-heating DM from its freeze-out to structure formation of the Universe.
In Sec.~\ref{sec:model_pion}, we discuss an extension of the SIMP model with a light ALP and dark radiation.
We conclude in Sec.~\ref{sec:conclusion}.
In Appendix~\ref{sec:thermal_history}, we provide a detailed derivation of the co-evolution equations of the DM number density and temperature.
In Appendix~\ref{sec:derivation_evolution_perturbations}, we present the evolution equations of cosmological perturbations in the self-heating DM scenario.

\section{Thermal history of self-heating DM from the freeze-out to structure formation}
\label{sec:self-heat_DM}
In this section, we discuss the freeze-out of the DM number density, its novel temperature evolution afterwards, and the evolution of cosmological perturbations in the self-heating DM scenario. 
For our purpose, we consider a scalar DM $\chi$ and a light mediator $\phi$. 
We mainly consider two types of interaction: One is self-scattering $\chi\chi \leftrightarrow \chi\chi$, and the other one is semi-annihilation $\chi\chi \leftrightarrow \chi \phi$, which are the minimal ingredients to realize self-heating of DM.
The presence of efficient self-scattering enforces the DM distribution to be $f_{\chi} = (n_{\chi} / n_{\chi}^{\rm eq}) \exp[ - E_{\chi} / T_{\chi}]$, where $n_{\chi}^{\rm eq} = (m_{\chi}^{2} T_{\chi} / 2 \pi^{2}) K_{2} (m_{\chi} / T_{\chi})$ and $K_{2}$ is the 2nd-order modified Bessel function of the second kind.
We also assume that a light mediator remains in thermal equilibrium by contacting either to the SM sector or to the dark sector and, hence, $f_{\phi} = \exp[ - E_{\phi}  / T_{\phi}]$, where $T_{\phi} = T_{\rm SM}$ or $T_{\phi} = T_{\rm DR}$, respectively.
Its number density is thus given by $n_{\phi}^{\rm eq} = (m_{\phi}^{2} T_{\phi} / 2 \pi^{2}) K_{2} (m_{\phi} / T_{\phi})$.
In general, annihilation $\chi \chi \leftrightarrow \phi \phi$ and also elastic scattering $\chi \phi \leftrightarrow \chi \phi$ may exist.

\subsection{Homogeneous and isotropic evolution}
The co-evolution equations of $n_{\chi}$ and $T_{\chi}$ are given, respectively, as
\begin{align}
\dot{n}_{\chi} + 3 H n_{\chi} =& 
-n_{\chi} \langle \semi \rangle_{T_{\chi} T_{\chi}}
\left[ n_{\chi} - {\cal J} (T_{\chi}, T_{\phi}) n_{\chi}^{\rm eq}(T_{\chi}) \right] \,,
\label{eq:num_den} \\
\dot{T}_{\chi} + 3 H T_{\chi} \left( \frac{T_{\chi}}{\sigma_E}\right)^{2}
=&
- \left( \frac{T_{\chi}}{\sigma_{E}}\right)^{2}
 \frac{n^{\rm eq}_{\phi}(T_{\chi})}{n_{\chi}^{\rm eq}(T_{\chi}) }
\langle \Delta E \invsemi \rangle_{T_{\chi}, T_{\phi} = T_{\chi}} 
\left[ 
n_{\chi} - n_{\chi}^{\rm eq}(T_{\chi}) {\cal K}(T_{\chi}, T_{\phi}) \right] \nonumber \\
& + 2 \gamma_{\chi \phi \rightarrow \chi \phi} (T_{\phi} - T_{\chi}) \,,
\label{eq:temp}
\end{align}
where $\sigma_{\rm semi}$  ($\sigma_{\rm inv}$) is the cross section for the $\chi\chi \rightarrow \chi \phi$ ($\chi\phi \rightarrow \chi \chi$) process.
See Appendix~\ref{sec:thermal_history} for the derivation.
We have added the elastic scattering term for completeness.
The expression of momentum exchange rate $\gamma_{\chi\phi \rightarrow \chi \phi}$ can be found in Ref.~\cite{Binder:2016pnr} and thus is not repeated here.
We remark that self-scattering does not contribute to these equations because it conserves the number and energy of DM particles.
The relic abundance of $\chi$ coincides with the observed one when
\bea
\label{eq:semixsection}
\langle \sigma_{\rm semi} \vrel \rangle |_{T_{\chi} = T_{\chi, \, {\rm fo}}} = (\sigma \vrel)_{\rm can} (T_{\phi} / T_{\rm SM} )_{\rm fo} \,,
\eea where $(\sigma \vrel)_{\rm can} \simeq (3 \times 10^{-26} \unit{cm^{3} / s})$ is a canonical cross section reproducing the observed relic density of thermal DM~\cite{Gondolo:1990dk}.
The canonical cross section is rescaled by the factor of $(T_{\phi} / T_{\rm SM})_{\rm fo}$ because $T_{\rm SM,\, fo} \simeq (m_{\chi} / 20) (T_{\rm SM} / T_{\phi})_{\rm fo}$. 

\begin{figure}
\includegraphics[width=0.6\linewidth]{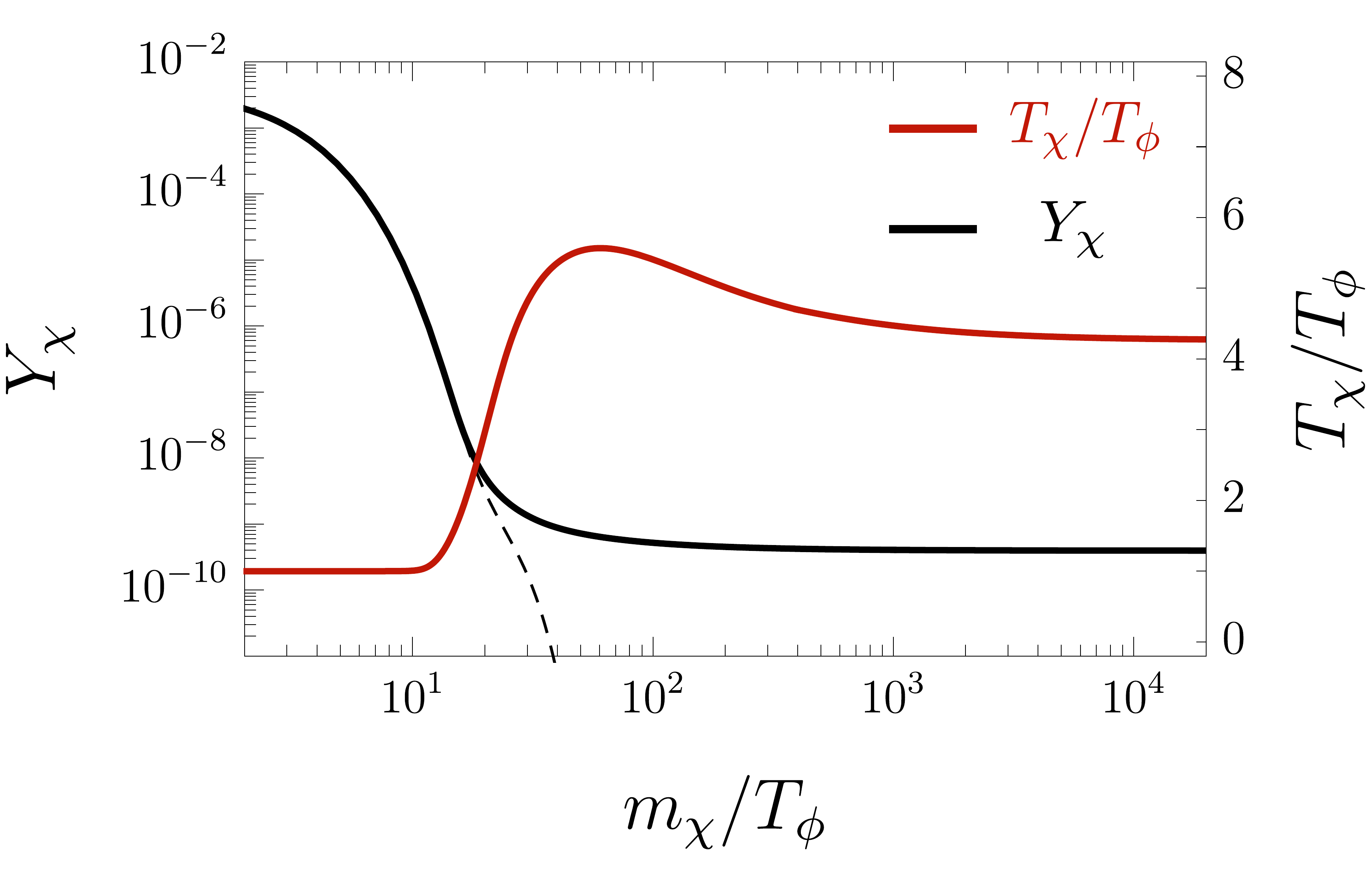}
\caption{%
Thermal history of self-heating DM. 
The plot shows the evolution of DM yield ({\bf black}) and also the evolution of the DM temperature ({\bf red}).
The dashed line corresponds to the equilibrium value.
For this plot, we assume no elastic scatterings; thus, the chemical and kinetic decoupling take place at the same time as semi-annihlation decouples from the plasma.
One can see that the DM temperature scales as $T_{\phi}$ scales even though DM is non-relativistic and kinetically decoupled from the thermal bath.}
\label{fig:thermal_history}
\end{figure}

We present numerical results in Fig.~\ref{fig:thermal_history}, which show the evolution of DM yield $Y_{\chi} = n_{\chi} / s$ and the temperature ratio $T_\chi/T_\phi$. 
Here $s = (2 \pi^{2} / 45) \, g_{*s, \, {\rm SM}} T^{3}$ is the entropy density, and $g_{*s,\, {\rm SM}}$ is the effective number of relativistic degrees of freedom in the SM sector.
In this figure, we assume that $\gamma_{\chi \phi \rightarrow \chi \phi} / H \ll 1$ and $T_{\phi} = T_{\rm SM}$.
Like usual thermal DM with $T_{\chi} = T_{\phi}$, the relic abundance is determined around $T_{\phi, \, {\rm fo}} \simeq m_{\chi} / 20$ (see Appendix~\ref{sec:thermal_history} for a small difference).
On the other hand, the DM temperature evolution shows a unique behavior.
Especially, after the freeze-out, the DM temperature scales as $T_{\chi} \propto 1/a$ (see Appendix~\ref{sec:thermal_history} for a thermodynamic derivation) despite the fact that DM particles are non-relativistic and that no elastic scattering equilibrates $T_{\chi}$ and $T_{\phi}$.
Self-heating of DM occurs because a small portion of DM still undergoes semi-annihilation after the freeze-out, and gain the kinetic energy of the order of its mass, which is much larger than the DM temperature.
We find that the ratio between the two temperatures approaches%
\footnote{%
The temperature ratio is constant up to changes in the number of relativistic degrees of freedom.}
\bea
r_{\chi \phi} \equiv \frac{T_{\chi}}{T_{\phi}} 
= (\gamma - 1)
\frac{2 m_{\chi}}{3 T_{\phi, \, {\rm fo}}} \,,
\eea
where $\gamma = (5/4)[ 1 - m_{\phi}^{2} / (5m_{\chi}^{2}) ]$ is the Lorentz boost factor of the final state DM particle.

\begin{figure}
\begin{tabular}{ccc}
\includegraphics[width=0.75\linewidth]{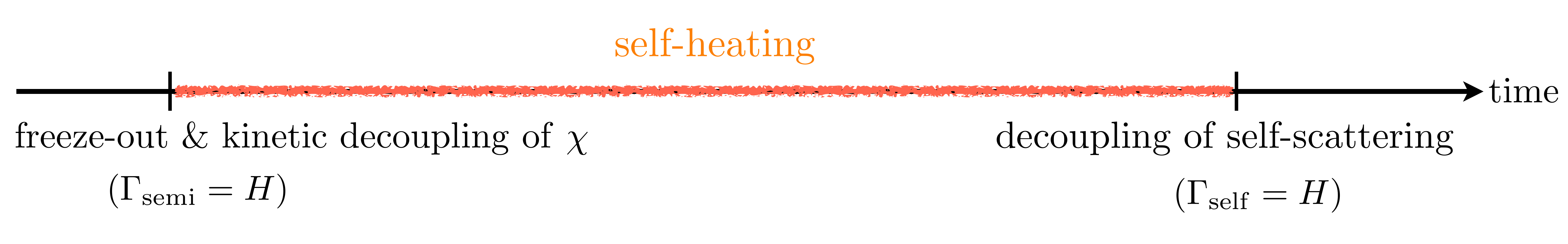}
\\
$\mathbf{\left(a\right)}$
\\
\\
\includegraphics[width=0.75\linewidth]{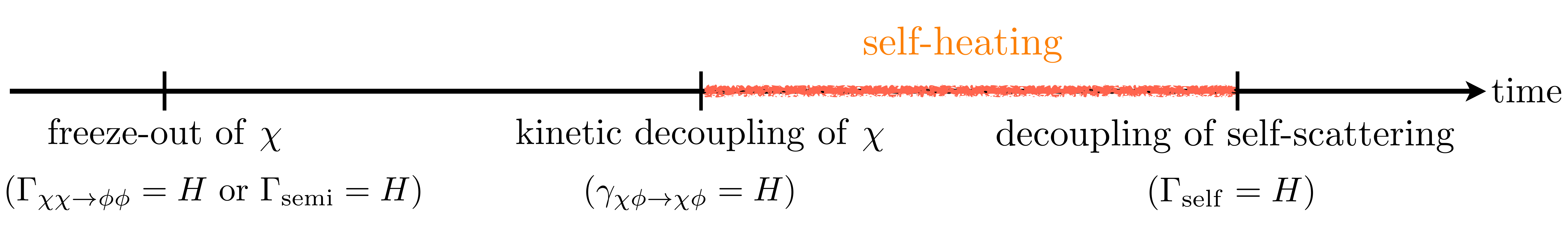}
\\
$\mathbf{\left(b\right)}$
\end{tabular}
\caption{%
The self-heating takes place between the kinetic decoupling and the freeze-out of DM self-interaction.
$({\bf a})$ $(\gamma_{\chi \phi \rightarrow \chi \phi} / H )|_{T_{\chi}=T_{\chi,\rm fo}} \ll 1$. 
The chemical and kinetic decoupling take place simultaneously. 
The self-heating of DM begins from the freeze-out of semi-annihilation to the freeze-out of self-interaction.
$({\bf b})$ $(\gamma_{\chi \phi \rightarrow \chi \phi} / H)|_{T_{\chi} = T_{\chi,\rm fo} } \gg 1$. 
The kinetic equilibrium could be maintained after the chemical freeze-out. 
In this case, the self-heating begins after the decoupling of elastic scattering, and continues until $\Gamma_{\rm self} = H$.
}
\label{fig:self_heating}
\end{figure}

Although we ignored elastic scatterings above, we emphasize that the self-heating is a generic feature of exothermic semi-annihilation even in the presence of elastic scattering.
If $(\gamma_{\chi \phi \rightarrow \chi \phi} / H)|_{T_{\chi} = T_{\chi, \, {\rm fo}} } \gg 1$, the elastic scattering is able to maintain the kinetic equilibrium of DM after the freeze-out, resulting in $T_{\chi} =T_{\phi}$. 
In this case, Eq.~\eqref{eq:num_den} reproduce a usual discussion found in Refs.~\cite{Hambye:2008bq,DEramo:2010keq, Belanger:2012vp}, since ${\cal J} (T_{\chi} =T_{\phi}, T_{\phi}) = 1$.
Eventually, elastic scattering decouples.
After this kinetic decoupling, DM begins to self-heat, and the DM temperature continues to scale as $T_{\chi} \propto 1/a$.
See Fig.~\ref{fig:self_heating} for schematic picture of the DM temperature evolution.

We stress that DM self-scattering is necessary for DM self-heating.
If self-scattering is absent, semi-annihilation merely produces a small portion of boosted DM particles, which act as a hot component of DM.
Thus, the unique temperature evolution continues only until $\Gamma_{\rm self} = \langle \self \rangle n_{\chi} \simeq H$. 
The larger the self-scattering cross section is, the longer the self-heating lasts and thus results in a larger thermal motion of DM.
The self-scattering decouples at
\begin{align}
\label{eq:selffreezeout}
T_{\rm SM, \, self} \simeq 
1 \eV  \, r_{\chi\phi}^{-n}
\left( \frac{1 \unit{cm^{2}/g}}{\sigma_{\rm self} / m_{\chi}} \right)^{2n} 
\left( \frac{m_{\chi}}{1 \GeV} \right)^{n} 
\left( \frac{T_{\rm SM}}{T_{\phi}} \right)^{n}_{\rm self} 
\, ,
\end{align}
where $n = 1/3$ for $T_{\rm SM, \, self} >  T_{\rm SM,\,eq}$, while $n=1/4$ when $T_{\rm SM, \, self} < T_{\rm SM,\,eq}$.
Here, $T_{\rm SM, \, eq} \simeq 0.8 \eV$ is the SM temperature at the matter-radiation equality.
When deriving Eq.~\eqref{eq:selffreezeout}, we have used $\left\langle v_{\rm rel} \right\rangle = (4/\sqrt{\pi}) \sqrt{T_{\rm SM}/m_\chi}\, r_{\chi\phi}^{1/2} (T_\phi/T_{\rm SM})^{1/2}$ and solve $\Gamma_{\rm self} = H$ for $T_{\rm SM}$. 
Remember that $\sigma_{\rm self} / m_{\chi} \sim 1 \unit{cm^{2}/g}$ forms a kpc core in a subgalactic halo~\cite{Tulin:2017ara}.%
\footnote{%
It is claimed that semi-annihilation cooperates with self-interaction to flatten the inner density profile of a subgalactic halo~\cite{Chu:2018nki}. 
The impact of the self-heating is more significant in smaller halos, and it may alleviate a required strength of self-scattering for solving the core-cusp problem.
We do not take this effect into account for simplicity in this paper.
}
A resultant large thermal motion of sub-GeV SIDM leaves a subgalactic-scale cutoff in the linear matter power spectrum like keV WDM.
We will discuss the structure formation of self-heating DM in the next section.

Before closing this section, we emphasize that the thermal history of $\phi$ plays a significant role both in DM searches and in the evolution of the Universe. 
If $\phi$ is massless, it changes the expansion rate of the Universe, as it contributes to the total energy density. 
Its impact is described by the change in the effective number of neutrino degrees of freedom $\Delta N_{\rm eff}$ and is constrained by big bang nucleosynthesis (BBN) and CMB.
The latest constraint from the {\it Planck} Collaboration is $\Delta N_{\rm eff} = 3.15 \pm 0.24$~\cite{Ade:2015xua}. 
For massive $\phi$, it may overclose the Universe unless it decays or they annihilate into light particles. 
If $\phi$ decays into visible particles, the late time production of $\phi$ through semi-annihilation is severely constrained by the Galactic and extra-Galactic gamma-ray searches as well as the CMB measurement.
These constraints require $m_{\chi} \gtrsim 10 \GeV$~\cite{Arcadi:2017kky, Roszkowski:2017nbc}, which results in a shorter duration of self-heating [see Eq.~\eqref{eq:selffreezeout}].%
\footnote{%
Thermal relic DM with a sub-GeV mass is still viable if DM particles annihilate into heavier particles~\cite{DAgnolo:2015ujb}.
In the case of semi-annihilation, this can be realized if $\phi$ is heavier than the DM mass~\cite{Kamada:2017tsq}.
In this case, we expect no self-heating of DM, because semi-annihilation is no longer exothermic.
} 
The other possibility is that $\phi$ decays into dark radiation.
In this case, constraints from DM indirect searches cannot be applied, although $\Delta N_{\rm eff}$ still constrains the scenario.
It will be discussed in Section~\ref{sec:model_pion}.

\subsection{Perturbed evolution}
The novel evolution of the DM temperature affects the evolution of cosmological perturbations. 
We derive evolution equations of cosmological perturbations in the self-heating scenario in Appendix~\ref{sec:derivation_evolution_perturbations}.
In the case of self-scattering DM, relevant variables are density contrast $\delta_{\chi}$, velocity divergence $\theta_{\chi}$, and entropy perturbation $\pi_{\chi}$.
The linearized equations for the cosmological perturbation for self-heating DM are given as
\begin{align}
\delta_{\chi}' =&
- \theta_{\chi} - 3 \Phi' \,,
\label{eq:delta_ev_syn} \\
\theta_{\chi}' =&
- {\cal H} \theta_{\chi} 
+ k^{2} \Psi + k^{2} (c_{s \chi}^{2} \delta_{\chi} + \pi_{\chi}) \,,
\label{eq:theta_ev_syn} \\
\pi_{\chi}' =& 
- 2 {\cal H} \pi_{\chi}
+ \left( c_{s \chi}^{2} - \frac{5}{3} \omega_{\chi} \right) \left[ - {\cal H} \left( 1 - \frac{{\cal H}'}{{\cal H}^{2}} \right) \delta_{\chi} + \theta_{\chi} + 3 ( \Phi' - {\cal H} \Psi) \right]
\label{eq:pi_ev_syn} \,,
\end{align}
where the prime is a derivative with respect to the conformal time and ${\cal H} = a' / a$. 
Here we take the conformal Newtonian gauge:
\begin{align}
ds^{2} =
a^{2} (\tau) 
\left[ - (1+ 2 \Psi) d \tau^{2} + ( 1 + 2 \Phi) \delta_{ij} dx^{i} dx^{j} \right] \,.
\label{eq:newtoniangauge}
\end{align}
We cross-check the result in the synchronous gauge in 
Appendix~\ref{sec:derivation_evolution_perturbations}.
Semi-annihilation affects $\delta_{\chi}$ mainly through the relatively large $c_{s \chi}^{2} \simeq (4 / 3) T_{\chi} / m_{\chi}$.
One can obtain the matter power spectrum by solving the above equations with equations of the equation of state $\omega_{\chi}$ and sound speed squared $c_{s \chi}^{2}$:
\begin{align}
\omega_{\chi}' &= 
-2 {\cal H} \omega_{\chi}
+ \frac{2}{3} (\gamma - 1) a \Gamma_{\rm semi} 
\left(1 - e^{-\Gamma_{\rm self} /  H} \right)\, ,
\label{eq:omega_eq_cut}
\\
c_{s \chi}^{2}  &= \frac{5}{3} \omega_{\chi}
-\frac{2}{9} (\gamma -1) \frac{a \Gamma_{\rm semi}}{\cal H} 
\left( 1 - e^{-\Gamma_{\rm self} / H } \right) \,,
\label{eq:cs_eq_cut}
\end{align}
with $\Gamma_{\rm semi} = n_{\chi} \langle \semi \rangle \propto 1 / a^{3}$.
As the interaction rate for self-scattering becomes smaller than the Hubble expansion rate, the DM temperature behaves as that of free-streaming non-relativistic particles, i.e., $T_{\chi} \propto 1/a^{2}$.

\begin{figure}
\centering
\begin{minipage}{0.45\linewidth}
\includegraphics[width=1.0\linewidth]{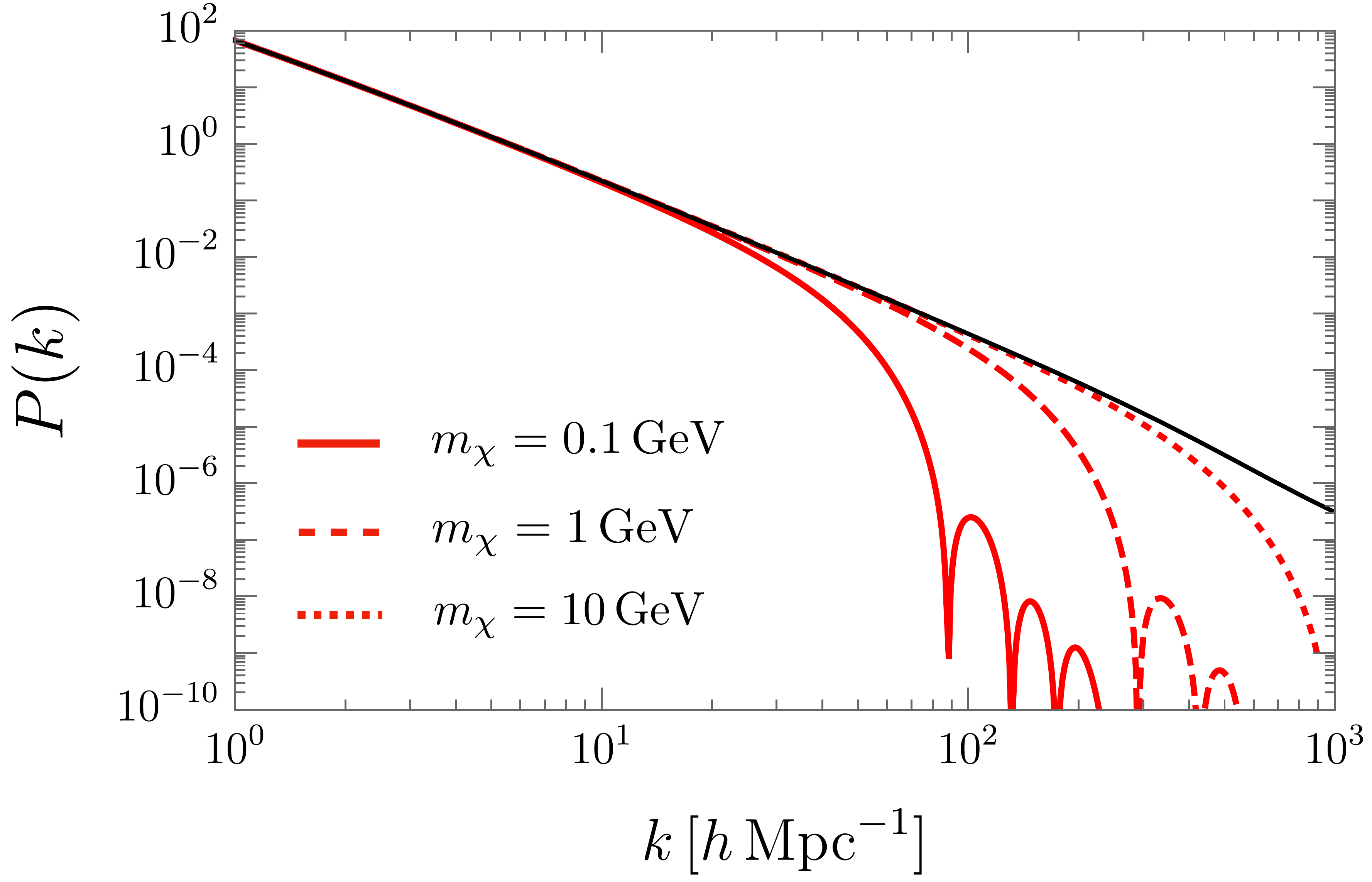}
\end{minipage}
\begin{minipage}{0.5\linewidth}
\includegraphics[width=1.0\linewidth]{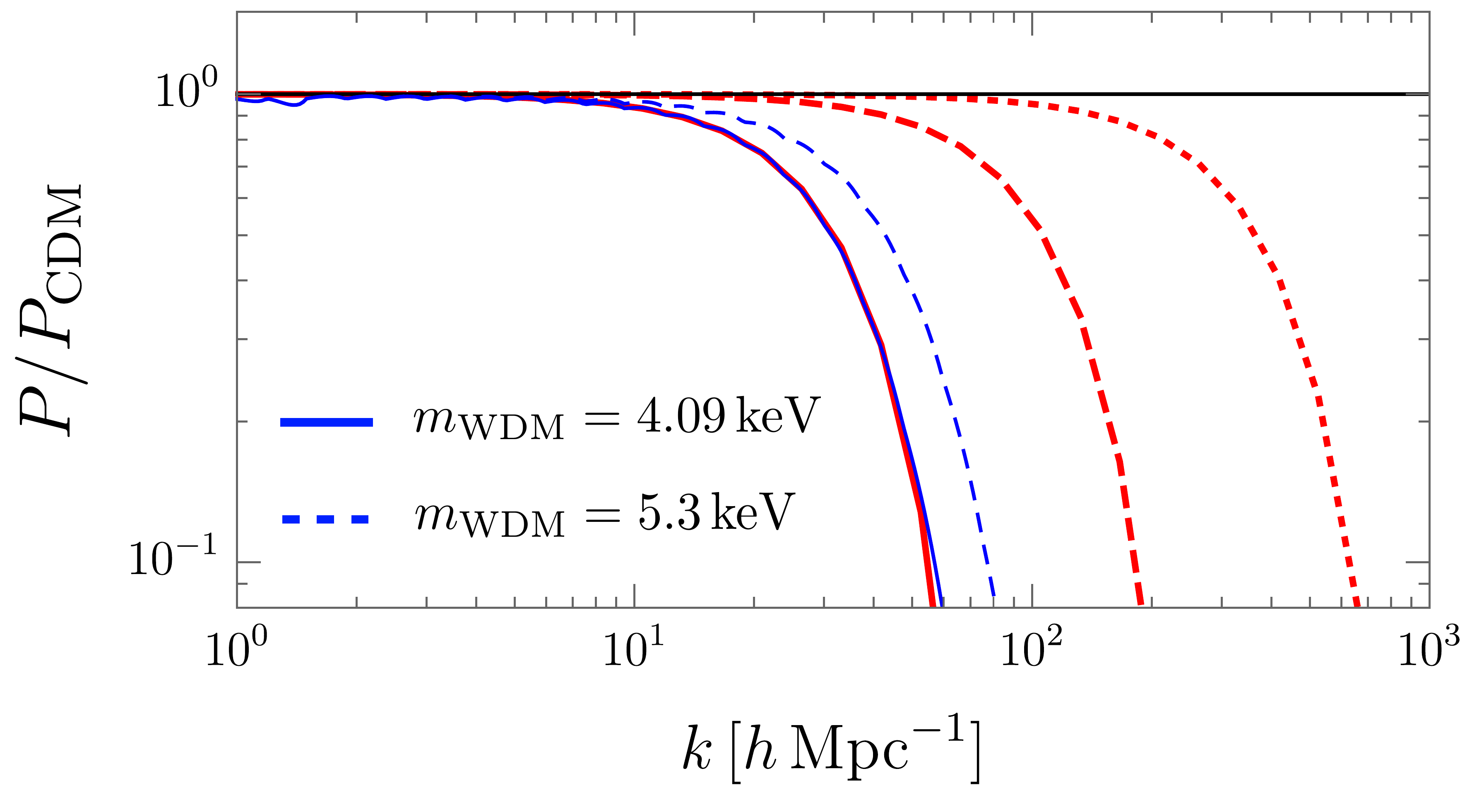}
\end{minipage}
\caption{%
(Left) Linear matter power spectrum for self-heating DM ({\bf red}) and CDM ({\bf black}) at the present Universe.
(Right) Linear matter power spectrum normalized with respect to that of CDM.
We also show the power spectrum in the thermal WDM models ({\bf blue}) with $m_{\rm WDM} = 4.09 \keV$ and $m_{\rm WDM} = 5.3 \keV$, where the latter is the latest lower bound on the mass of WDM~\cite{Irsic:2017ixq}.
We choose $\sigma_{\rm self} / m_{\chi} = 1 \unit{cm^{2} / g}$ such that self-scattering decouples at $T_{\rm SM, \, self} = 0.5 \eV, \, 0.8 \eV,$ and $1.6 \eV$ for $m_{\chi} = 0.1 \GeV,\, 1 \GeV,$ and $10 \GeV$, respectively.
We also assume $T_{\phi} = T_{\rm SM}$.}
\label{fig:matterPS}
\end{figure}

By modifying the publicly available Boltzmann solver {\tt CLASS}~\cite{Blas:2011rf}, we obtain the present linear matter power spectrum as shown in Fig.~\ref{fig:matterPS}.
The matter power spectrum exhibits a sharp cutoff around $k = {\cal O}(100) \unit{Mpc^{-1}}$.
This scale turns out to well match the Jeans instability scale at the matter-radiation equality~\cite{Kamada:2013sh, Kamada:2017gfc}:
\begin{align}
k_{\rm J} \simeq 
180 \, {\rm Mpc}^{-1} \, r_{\chi\phi}^{-1/2}
\, \max\left(1, \sqrt{a_{\rm eq} / a_{\rm self}} \right)
\left( \frac{m_{\chi} }{1 \GeV} \right)^{1/2}
\left( \frac{T_{\gamma}}{T_{\phi}} \right)_{\rm eq}^{1/2} \,,
\end{align}
where the temperature ratio $r_{\chi \phi} = T_{\chi} / T_{\phi}$ should be evaluated at $a = \min(a_{\rm self},\, a_{\rm eq})$. 

We compare the resultant matter power spectra in self-heating DM to those in the thermal WDM model (see, e.g., Ref.~\cite{Bae:2017dpt} for details).
The Lyman-$\alpha$ forest data constrain the WDM mass as $m_{\rm WDM} > 5.3 \keV$~\cite{Irsic:2017ixq}. 
In the case of WDM, the Jeans instability scale appears to be $k_{\rm J, WDM} \simeq 180 \unit{Mpc^{-1}} \left(m_{\rm WDM} / 5.3 \, {\rm keV} \right)^{4/3}$, while a drop of power in the matter power spectrum takes place around $k_{\rm J, WDM} / 4$, where an additional order one factor can be attributed to the free-streaming of WDM during the radiation-dominated Universe~\cite{Kamada:2017icv}.
By equating $k_{\rm J} \simeq k_{\rm J, WDM} /4$, we find the following correspondence between WDM and self-heating DM:
\begin{align}
\frac{m_{\rm WDM}}{ 5.3\keV} \simeq
\left( \frac{r_{\chi\phi}}{2.4} \right)^{-3/8}
\left(\frac{m_{\chi}}{0.1 \GeV} \right)^{3/8} 
\max \left(1, \sqrt{a_{\rm eq} / a_{\rm self}} \right)^{3/4}
\left( \frac{T_{\gamma}}{T_{\phi}} \right)_{\rm eq}^{3/8}
\,.
\end{align}
For the massless mediator $\phi$ sharing the temperature with SM particles, the current limit on the thermal WDM mass, $m_{\rm WDM} \geq 5.3 \keV$, translates into the mass bound of self-heating DM as $m_{\chi} \geq 0.1 \GeV$ if the self-interaction decouples after the matter-radiation equality.

\section{Extended SIMP model with an ALP and dark radiation}
\label{sec:model_pion}
In this section, we propose a realization of self-heating DM.
We extend the SIMP model~\cite{Hochberg:2014kqa} with an ALP~\cite{Kamada:2017tsq} by introducing dark radiation.
The Lagrangian density is given by
\begin{align}
\label{eq:uvlagran}
{\cal L} =& \left| \partial_{\mu} \Phi \right|^{2} - V (|\Phi|^{2}) + N^{\dagger} i \bar{\sigma}^{\mu} D_{\mu} N + \bar{N}^{\dagger} i \bar{\sigma}^{\mu} D_{\mu} \bar{N} - \left(m_{N} \bar{N} N + {\rm h.c.} \right) \nonumber \\
& + Q^{\dagger} i \bar{\sigma}^{\mu} D_{\mu} Q + \bar{Q}^{\dagger} i \bar{\sigma}^{\mu} D_{\mu} \bar{Q} + L^{\dagger} i \bar{\sigma}^{\mu} D_{\mu} L + \bar{L}^{\dagger} i \bar{\sigma}^{\mu} D_{\mu} \bar{L} - \Phi \left( y_{Q} Q \bar{Q} + y_{L} L \bar{L} + {\rm h.c.} \right) \nonumber \\
& - \frac{1}{4} H_{~ \mu \nu}^{a} H^{a \mu \nu} - \frac{1}{4} X_{~ \mu \nu}^{a} X^{a \mu \nu} + \theta_{H} \frac{g_{H}^{2}}{32 \pi^{2}} H_{~ \mu \nu}^{a} {\widetilde H}^{a \mu \nu} + \theta_{X} \frac{g_{X}^{2}}{32 \pi^{2}} X_{~ \mu \nu}^{a} {\widetilde X}^{a \mu \nu} \,,
\end{align}
where $H_{~ \mu \nu}^{a}$ (${\widetilde H}_{~ \mu \nu}^{a}$) is the (dual) field strength of the confining SU$(N_{H})$ gauge field and $X_{~ \mu \nu}^{a}$ (${\widetilde X}_{~ \mu \nu}^{a}$) is the (dual) field strength of the quasi-perturbative SU$(N_{X})$ gauge field.
Gauge charges of the matter contents are summarized in Table~\ref{tab:charges}.
$N$ and $\bar{N}$ are $N_{f}$-flavored ($N_{f} \geq 3 $).

\begin{table}
\centering
\begin{tabular}{c|ccc}
& SU(N$_{H}$) & SU(N$_{X}$) & U(1)$_{\rm PQ}$\\ \hline
$Q$          & $\square$  & 1 & $-1/2$  \\
$\bar{Q}$ &  $\bar{\square}$  & 1 & $-1/2$\\
$L$          & 1 & $\square$ & $-1/2$  \\
$\bar{L}$ & 1& $\bar{\square}$ & $-1/2$   \\
$N$         & $\square$ & 1 & 0 \\
$\bar{N}$ & $\bar{\square}$ & 1 & 0 \\
$\Phi$     & 1 & 1& 1
\end{tabular}
\caption{Gauge charges of matter contents.}
\label{tab:charges}
\end{table}

As the PQ symmetry is spontaneously broken, an ALP arises. 
The mass of $Q$ and $L$ originates from vacuum expectation value $\langle\Phi\rangle$ and is assumed to be heavier than the confinement scale of the SU$(N_{H})$ gauge group.
After integrating out $Q$ and $L$, we find that the ALP couples to both $H_{~ \mu \nu}^{a} {\widetilde H}^{a \mu \nu}$ and $X_{~ \mu \nu}^{a} {\widetilde X}^{a \mu \nu}$.
Once SU$(N_{H})$ confines, SU$(N_{f})_{L} \times$SU$(N_{f})_{R}$ global symmetry breaks down to the diagonal subgroup SU$(N_{f})_{V}$, and pion-like particles $\chi$ emerge.
For low energy phenomenology, we obtain the following effective Lagrangian:
\begin{align}
\label{eq:anomalous}
{\cal L}_{\rm eff} = 
\frac{1}{2} \left(\partial_{\mu} \phi \right)^{2} 
- \frac{1}{4} X_{~ \mu \nu}^{a} X^{a \mu \nu}
+ \frac{g_{X}^{2}}{32\pi^{2}}  \frac{\phi}{f} X_{~ \mu \nu}^{a} {\widetilde X}^{a \mu \nu} 
+ {\cal L}_{\rm chiral} \,.
\end{align}
The expression of ${\cal L}_{\rm chiral}$ can be found in Ref.~\cite{Kamada:2017tsq} and thus is not repeated here.
We emphasize that SU$(N_{f})_{V}$ is the exact symmetry of the theory; hence, no decay operator of DM pions is allowed.
The ALP $\phi$ and dark radiation $X$ form a dark plasma with the temperature $T_{\phi} = T_{\rm DR}$. 
In the following, we take $N_{f}=4$, $N_{H}=3$, and $N_{X}=2$ as a benchmark for the analysis.

Low energy phenomenology is described by three parameters: DM pion mass $m_{\chi}$, DM pion decay constant $f_{\chi}$, and ALP decay constant $f$. 
The self-scattering cross section for solving the core-cusp problem, and the semi-annihilation cross section for the observed DM relic density [see Eq.~\eqref{eq:semixsection}] are achieved as~\cite{Kamada:2017tsq}
\begin{align}
\label{eq:pidecayconst}
\frac{\sigma_{\rm self}}{m_{\chi}} \, =& \,
1 \unit{cm^{2} / g}
\left( \frac{m_{\chi}}{50\MeV} \right) \left( \frac{40 \MeV}{f_{\chi}} \right)^{4}
\,, \\
\langle \semi \rangle \, =& \,
(\sigma \vrel)_{\rm can}
\left( \frac{m_{\chi}}{50\MeV} \right)^{2} 
\left( \frac{40\MeV}{f_{\chi}} \right)^{2}
\left( \frac{300\GeV}{f} \right)^{2} \,.
\end{align}
Note that we only consider CP-conserving interactions of DM pions and ALPs, because the CP-violaing vacuum angle in the hidden confining sector dynamically vanishes due to an ALP. 
Although we consider semi-annihilation as a dominant process for the freeze-out of the DM number density, there exist two additional number-changing processes: the 3-to-2 process of $\chi \chi \chi \rightarrow \chi \chi$ and annihilation $\chi \chi \rightarrow \phi \phi$. 
Since the annihilation cross section is suppressed by $(f_{\chi} / f)^{2}$ compared to that of semi-annihilation, it is sub-domnant.
Meanwhile, the 3-to-2 process is sub-dominant for
\begin{align}
m_{\chi} < 230\MeV\left(\frac{T_{{\rm DR}}}{T_{{\rm SM}}}\right)_{{\rm fo}}^{8/9},
\end{align}
where we have chosen $f_{\chi}$ and $f$ such that $\sigma_{\rm self}/m_{\chi} \simeq 1 \unit{cm^{2} / g}$ and $\langle \semi \rangle = (\sigma \vrel)_{\rm can} (T_{\rm DR} / T_{\rm SM})_{\rm fo}$.
Note that the ALP obtains its mass from explicit breaking of chiral symmetry of the QCD-like sector, and its mass is given as~\cite{Kamada:2017tsq}
\begin{align}
\label{eq:mphiscaling}
m_{\phi} 
= \frac{m_{\chi} f_{\chi}}{2 \sqrt{2 N_{f}} f} 
\simeq 
1 \keV \, 
\left( \frac{m_{\chi}}{50\MeV} \right)
\left( \frac{f_{\chi}}{40\MeV} \right)
\left( \frac{300\GeV}{f} \right).
\end{align}

The elastic scattering between the ALP and DM also exists in this model. 
For the self-heating to occur, the elastic scattering should decouple before the self-scattering decouples.
This can be trivially achieved in our model, because the strength of elastic scattering is suppressed by $(f_{\chi} / f)^{2}$ compared to that of semi-annihilation, and by $(f_{\chi}/f)^4$ compared to that of self-interaction.
As a consequence, the momentum exchange rate due to elastic scattering is
\begin{align}
\frac{\gamma_{\chi\phi\rightarrow\chi\phi}}{H}\simeq 
5 \times10^{-4}  \,\, r_{\chi\phi}^{-1}
\left( \frac{300 \GeV}{f} \right)^{4} 
\left( \frac{m_{\chi}}{50 \MeV} \right)
\left( \frac{10.75}{g_{*,\,{\rm SM}} (T_{\rm SM})} \right)^{1/2} 
\left( \frac{T_{\rm DR}}{T_{\rm SM}} \right)^{4} 
\left( \frac{T_{\rm SM}}{1\MeV} \right)^{2}.
\end{align}
Thus, elastic scattering is inefficient during and after the freeze-out of DM, and the self-heating begins roughly after the freeze-out of DM.

We assumed that the ALP and dark gauge boson form a thermal bath with $T_{\rm DR}$ during the freeze-out of DM.
The $X X\rightarrow X X X$ rate~\cite{Biro:1993qt, Xu:2007ns} is sufficiently large for $\alpha_{X} > {\cal O} (10^{-10})$. 
The $\phi X \rightarrow XX$ keeps $\phi$ in the thermal bath of $X$ until $T_{\rm SM} = 1\MeV$ as~\cite{Masso:2002np,Graf:2010tv}
\begin{align}
\frac{\Gamma_{\phi X \rightarrow X X}}{H} \simeq&\, 
4\times 10^{3} 
\left( \frac{\alpha_{X}}{10^{-2}} \right)^{3} 
\left( \frac{300 \GeV}{f} \right)^{2} 
\left( \frac{10.75}{g_{*, \,{\rm SM}} (T_{\rm SM})} \right)^{1/2} \left( \frac{T_{\rm DR}}{T_{\rm SM}} \right)^{3} \left( \frac{T_{\rm SM}}{1\MeV} \right) \,.
\end{align}
An ALP decays when it is still semi-relativistic with the thermally averaged decay rate of
\begin{align}
\frac{\left\langle \Gamma_{\phi\rightarrow XX}\right\rangle }{H}\simeq 
0.4
\left(\frac{\alpha_{X}}{10^{-2}}\right)^{2}
\left(\frac{m_{\phi}}{1\keV}\right)^{4}
\left(\frac{300\GeV}{f}\right)^{2} 
\left(\frac{10.75}{g_{*,\,{\rm SM}}(T_{{\rm SM}})}\right)^{1/2}
\left(\frac{T_{{\rm DR}}}{T_{{\rm SM}}}\right)^{-1}
\left(\frac{1\keV}{T_{{\rm SM}}}\right)^{3} \,.
\end{align}
Non-Abelian dark radiation does not confine until the present Universe for $\alpha_{X} (T_{\rm DR} = T_{\rm DR, \, fo}) \lesssim 0.03\, (2 / N_{X})$, since the confinement scale is given by $\Lambda = \mu_{0} \exp\left[- 6 \pi  / (11 N_{X} \alpha_{X,0}) \right]$, where $\alpha_{X,0} = \alpha_{X} (T_{\rm DR} = \mu_{0})$.
Neither the ALP nor dark radiation overcloses the Universe.

The model Lagrangian~\eqref{eq:uvlagran} does not contain any interaction that equilibrates the SM and dark sector.
Indeed, if two sectors are in thermal equilibrium with each other in the early Universe, $\Delta N_{\rm eff}$ tends to exceed unity, which is strongly disfavored by BBN and CMB.  
The inflaton can decay both into the SM sector and into the dark sector.
In this case, $(T_{\rm DM} / T_{\rm SM})$ is determined by the branching ratio.
Even if the inflaton predominantly decays into the SM sector, the dark sector can be populated through a feeble interaction to the SM sector, while not being completely thermalized with SM particles.
In addition to the interactions in Eq.~\eqref{eq:uvlagran}, we may consider a Higgs portal coupling:
\begin{align}
{\cal L}_{H \Phi} = \lambda_{H \Phi} |\Phi|^{2} |H|^{2} \,.
\end{align}
The continuous production of the dark sector increases the temperature ratio between the dark sector and the SM plasma until the electroweak phase transition.
The temperature ratio between the two sectors at the electroweak phase transition is estimated as
\begin{align}
\left( \frac{T_{\rm DR}}{T_{\rm SM}} \right)_{\rm ew}
=\left( \frac{g_{*, \, {\rm SM}} \, \rho_{\rm DR}}{g_{*, \, {\rm DR}} \, \rho_{\rm SM}} \right)_{\rm ew}^{1/4}
\simeq 0.5\left(\frac{\lambda_{H\Phi}}{2.2\times10^{-6}}\right)^{1/2}\left(\frac{106.75}{g_{*,\,{\rm SM, \,ew}}}\right)^{1/8}\left(\frac{83.5}{g_{*,\,{\rm DR, \,ew}}}\right)^{1/4} \,,
\end{align}
where $g_{*, \, {\rm DR, \, ew}} = 83.5$ takes into account all the degrees of freedom of particles in Eq.~\eqref{eq:uvlagran}.
The contribution of the dark sector to $\Delta N_{\rm eff}$ at the neutrino decoupling is given by
\begin{align}
\Delta N_{\rm eff} = \frac{8}{7} \times \frac{1}{2} \times g_{* s, \, {\rm DR}} \times \left( \frac{T_{\rm DR}}{T_{{\rm SM}}} \right)^{4}_{\text{$\nu$-dec}}
\simeq9.8\,g_{*s,\,{\rm DR}}^{-1/3}\left(\frac{g_{*s,\,{\rm DR,\,ew}}}{83.5}\right)^{4/3}\left(\frac{T_{{\rm DR}}}{T_{{\rm SM}}}\right)_{{\rm ew}}^{4} \,.
\end{align}
After the freeze-out of DM but before the decay of the ALP, we find $g_{*s,\, {\rm DR}} = 7$.
Thus, $(T_{\rm DR}/T_{\rm SM})|_{\rm ew} \simeq 0.5$ is consistent with $\Delta N_{\rm eff} < 3.4$~\cite{Ade:2015xua}.

\section{Conclusion}
\label{sec:conclusion}
Self-heating of semi-annihilating DM can suppress subgalactic-scale structure formation when it lasts until the matter-radiation equality. 
The reduced number of dwarf-size halos can reconcile the possible tension between the CDM paradigm and the observation.
The self-heating of sub-GeV DM is maintained until the matter-radiation equality for $\sigma_{\rm self}/m_{\chi}\simeq {\cal O}(0.1\text{--}1) \unit{cm^{2} / g}$.
We have followed the evolution of cosmological perturbations and demonstrated that self-heating sub-GeV DM indeed leaves a cutoff  on the subgalactic scale of the linear matter power spectrum.

It is interesting that self-heating DM interrelates a subgalactic cutoff in the linear matter power spectrum and a kpc core of the DM distribution in halos through the thermalization of DM particles.
We can take full advantage of astrophysical and cosmological searches of WDM and SIDM to probe self-heating DM.
For example, we could tighten the range of the self-interaction strength by analyzing line-of-sight velocity dispersions of dwarf spheroidal galaxies~\cite{Valli:2017ktb} and rotation curves of low-surface brightness galaxies~\cite{Bondarenko:2017rfu} with a larger number of samples.
The satellite number counts restrict the cutoff in the linear matter power spectrum~\cite{Polisensky:2010rw, Kim:2017iwr}.
The matter distribution smoother than the CDM prediction will be tested by the perturbations on strongly lensed systems~\cite{Inoue:2014jka, Kamada:2016vsc, Birrer:2017rpp, Gilman:2017voy}.
The top-down structure formation in contrast to the bottom-up one in the CDM paradigm is tested by the further discoveries of high-$z$ galaxies~\cite{Barkana:2001gr, Pacucci:2013jfa, Lovell:2017eec} and by multiple probes of the reionization epoch such as the $21 \unit{cm}$ brightness temperature~\cite{Sitwell:2013fpa,Safarzadeh:2018hhg,Schneider:2018xba} and its fluctuations due to minihalos~\cite{Sekiguchi:2014wfa}.

We have proposed an extension of the SIMP model with an ALP and dark radiation as a particle physics realization of sub-GeV self-heating DM.
DM pions semi-annihilate into an ALP, which decays into dark radiation.
Dark radiation and an ALP forms thermal equilibrium.
We have shown that self-heating can be realized for a certain range of model parameters.
When the dark sector is populated from the SM sector through a Higgs portal, we can produce the dark sector particles while being consistent with the constraints from BBN and CMB measurements.

We have focused on a velocity-independent self-scattering cross section.
On the other hand, the self-scattering cross section diminishing with an increasing velocity may be favored by the constraints from galaxy cluster ellipticities and bullet clusters.
One way to realize the velocity-dependent self-scattering cross section is to introduce a light mediator coupling to two DM particles.
Extending our discussion to such a case will be intriguing.

\begin{acknowledgements}
We would like to thank Toyokazu Sekiguchi for useful discussions in the early stage of this work.
We also thank Andrew Spray for carefully reading our manuscript.
This work was supported by IBS under the project code, IBS-R018-D1.
\end{acknowledgements}

\newpage
\appendix
\section{Evolution equations of the self-heating DM number density and temperature}
\label{sec:thermal_history}
The evolution of the DM phase-space distribution is governed by the Boltzmann equation,
\begin{align}
E_{\chi}
\left[ 
\frac{\partial f_{\chi}}{\partial t} - H p_{\chi} \frac{\partial f_{\chi}}{\partial p_{\chi}} 
\right]
= C [f_{\chi}] \,,
\label{eq:Boltzmann_leading}
\end{align}
with 
\begin{align}
C_{\rm semi} [f_{\chi}] =&
\int d\Pi_{2} d\Pi_{3} d\Pi_{4} \, 
(2\pi)^{4} \delta^{(4)} \left( p_{1} + p_{2} - p_{3} - p_{4} \right) |{\cal M}_{\rm semi}|^{2} 
\left[ f_{\chi}(p_{3}) f_{\phi}(p_{4}) - f_{\chi}(p_{1}) f_{\chi}(p_{2}) \right]
\nonumber \\
& + 
\frac{1}{4} \int d\Pi_{2} d\Pi_{3} d\Pi_{4} \, 
(2\pi)^{4} \delta^{(4)} \left( p_{1} + p_{2} - p_{3} - p_{4} \right) |{\cal M}_{\rm semi}|^{2} 
\left[ f_{\chi}(p_{3}) f_{\chi}(p_{4}) - f_{\chi}(p_{1}) f_{\phi}(p_{2}) \right] \,,
\label{eq:semi_collision}
\\
C_{\rm self} [f_{\chi}] =&
\frac{1}{2}
\int d\Pi_{2} d\Pi_{3} d\Pi_{4} \, 
(2\pi)^{4} \delta^{(4)} \left( p_{1} + p_{2} - p_{3} - p_{4} \right) |{\cal M}_{\rm self}|^{2} 
\left[ f_{\chi}(p_{3}) f_{\chi}(p_{4}) - f_{\chi}(p_{1}) f_{\chi}(p_{2}) \right] \,.
\end{align}
$|{\cal M}_{\rm semi}|^{2}$ and $d\Pi = d^{3} \vec{p} / (2 \pi)^{3} / 2 E$ denote the invariant amplitude squared and the invariant phase space measure, respectively.
Symmetry factors are multiplied in each integral in order not to overcount the phase space of identical particles.

We derive the evolution equations of the DM number density $n_{\chi}$ and temperature $T_{\chi}$ from the number and energy conservation equations.
We remark that self-scattering does not contribute to these equations because it conserves the number and energy of DM particles.
Integrating the Boltzmann equation with $1 / E_{\chi}$ over the phase space of DM, we find the evolution equation of $n_{\chi}$ as
\begin{align}
\dot{n}_{\chi} + 3 H n_{\chi} &= 
\frac{1}{2} \int \prod_{i=1}^{4} d \Pi_{i}
(2\pi)^{4} \delta^{(4)}(p_{1} + p_{2} - p_{3} -p_{4})|{\cal M}_{\rm semi}|^{2} 
\left[ f_{\chi}(p_{3}) f_{\phi}(p_{4}) - f_{\chi} (p_{1}) f_{\chi} (p_{2}) \right],
\nonumber \\
&= - \langle \semi \rangle_{T_{\chi} T_{\chi}} 
\left[ n_{\chi}^{2} - n_\chi n^{\rm eq}_{\chi}(T_\chi) {\cal J}(T_{\chi}, T_{\phi}) \right] \,,
\label{eq:detail_num}
\end{align}
where ${\cal J}(T_{\chi},T_{\phi})$ is defined as 
\begin{align}
{\cal J} (T_{\chi} , T_{\phi}) &= 
\frac{n_{\phi}^{\rm eq}(T_{\phi})}{n_{\phi}^{\rm eq}(T_{\chi})}
\frac{\langle \invsemi \rangle_{T_{\chi}, T_{\phi}} }{\langle \invsemi \rangle_{T_{\chi}, T_{\phi} =T_{\chi}}} 
\,.
\end{align}
In this derivation, we have used $\left( n_{\chi}^{\rm eq} \right)^{2} \langle \semi \rangle_{T_{\chi} T_{\chi}} = n_{\chi}^{\rm eq} n_{\phi}^{\rm eq} \langle \invsemi \rangle_{T_{\chi} T_{\chi}}$.
The subscript in $\langle \semi \rangle_{T_{\chi} T_{\chi}}$ denotes that this thermal average is defined as a thermal average with respect to the Boltzmann distribution with temperature $T_{\chi}$.
Because of the explicit dependence on $T_{\chi}$, the equation for the number density is not closed by itself.
To correctly describe the evolution of the system, one must track the DM temperature evolution as well.

To obtain the evolution equation of the DM temperature, we integrate the Boltzmann equation without any weight.
We find the evolution equation of the energy density:
\begin{align}
\dot{\rho}_{\chi} + 3 H ( \rho_{\chi} + P_{\chi}) 
=& \frac{1}{2} \int \prod_{i=1}^{4} d \Pi_{i} \, (E_{1}+E_{2} - E_{3}) (2\pi)^{4} \delta^{(4)} \left( p_{1} + p_{2} - p_{3} - p_{4} \right) |{\cal M}_{\rm semi}|^{2} \nonumber \\
& \times \left[ f_{\chi}(p_{3}) f_{\phi}(p_{4}) - f_{\chi}(p_{1})f_{\chi}(p_{2}) \right] \,,
\end{align}
with pressure $P_{\chi}$.
Since $\rho_{\chi} = \langle E_{\chi} \rangle n_{\chi}$ and $P_{\chi} = \left\langle p_{\chi}^{2} / \left( 3 E_{\chi} \right) \right\rangle n_{\chi} = T_{\chi} n_{\chi}$,
the above equation reads as
\begin{align}
\dot{\left\langle E_{\chi} \right\rangle}
+ 3 H T_{\chi} =& 
- \frac{n^{\rm eq}_{\phi}(T_{\chi})}{n_{\chi}^{\rm eq}(T_{\chi})}
\langle \Delta E \invsemi \rangle_{T_{\chi} T_{\chi}}
\left[ n_{\chi} - n_{\chi}^{\rm eq} (T_{\chi}) {\cal K}(T_{\chi}, T_{\phi})  \right] \,, 
\end{align}
where $\Delta E = E_{\phi} - \left\langle E_{\chi} \right\rangle_{T_{\chi}}$ and the function ${\cal K}(T_{\chi},T_{\phi})$ is defined as 
\begin{align}
{\cal K}(T_{\chi}, T_{\phi}) &= 
\frac{n^{\rm eq}_{\phi}(T_{\phi})}{n^{\rm eq}_{\phi}(T_{\chi})}
\frac{\langle \Delta E \invsemi \rangle_{T_{\chi}, T_{\phi}}}{\langle \Delta E \invsemi \rangle_{T_{\chi}, T_{\phi} = T_{\chi}}}
\, .
\end{align}
Using $\dot{\left\langle E_{\chi} \right\rangle} = ( \dot{T}_{\chi} / T_{\chi}^{2} ) ( \langle E_{\chi}^{2} \rangle - \langle E_{\chi} \rangle^{2} ) = ( \dot{T}_{\chi} / T_{\chi}^{2} ) \sigma_E^{2}$,
one obtains
\begin{align}
\frac{\dot{T}_{\chi}}{T_{\chi}^{2}} + 3 H \frac{T_{\chi}}{\sigma_E^{2}} 
= - \frac{1}{\sigma_E^{2}} 
\frac{n_{\phi}^{\rm eq} (T_{\chi})}{n_{\chi}^{\rm eq}(T_{\chi})}
\langle \Delta E \invsemi \rangle_{T_{\chi} T_{\chi}}
\left[
n_{\chi} -
n_{\chi}^{\rm eq} (T_{\chi}) {\cal K}(T_{\chi}, T_{\phi})
\right] \,.
\end{align}

\begin{figure}
\centering
\includegraphics[width=0.6\linewidth]{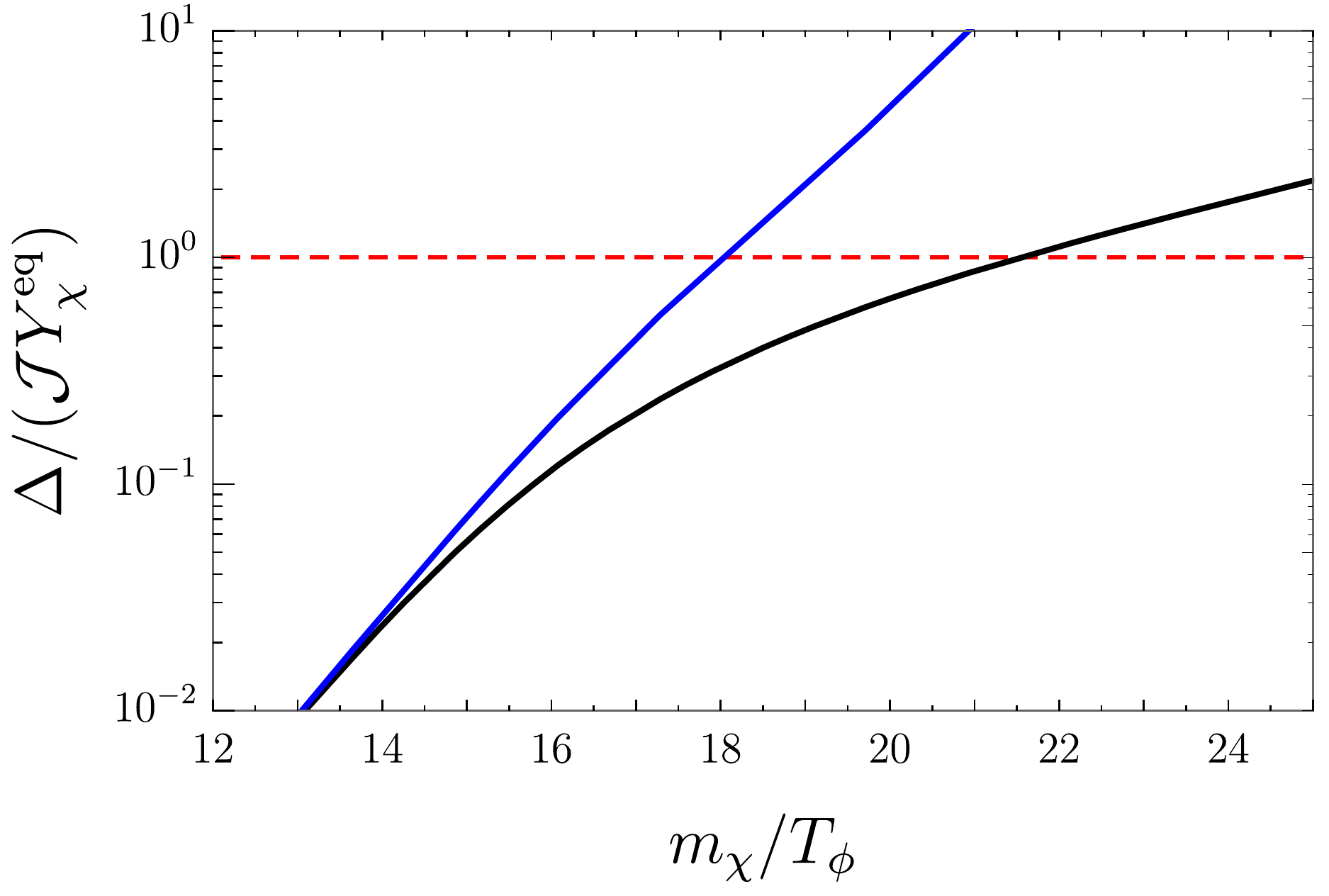}
\caption{$\Delta / ({\cal J}Y_{\chi}^{\rm eq})$ in the self-heating DM scenario ({\bf black}). 
Here we assume $T_{\phi} = T_{\rm SM}$ for simplicity.
Compared to the $T_{\chi} = T_{\rm SM}$ case ({\bf blue}), the freeze-out is delayed. }
\label{fig:delay_fo}
\end{figure}

Figure.~\ref{fig:thermal_history} presents our numerical result of the co-evolution of $n_{\chi}$ and $T_{\chi}$.
One can see that the freeze-out of DM yield $Y_{\chi}$ proceeds in a similar way to a usual discussion found in Refs.~\cite{Hambye:2008bq,DEramo:2010keq, Belanger:2012vp}.
The DM yield is estimated as
\bea
Y_{\chi, \infty} \simeq 
\left. \frac{H}{s \langle \semi \rangle} \right|_{T_{\phi} = T_{\phi, \, {\rm fo}}} \,,
\eea
while determining the freeze-out temperature is ambiguous in our case.
In the case of $T_{\chi} = T_{\phi}$, the freeze-out temperature is usually determined by $\Delta (x_{\rm fo})= Y_{\chi} - Y_{\chi}^{\rm eq} = c Y^{\rm eq}_{\chi} (x_{\rm fo})$ with $c$ being a numerical constant of order unity, where $\Delta = (d \ln Y^{\rm eq}_{\chi} / dx) / (- s \langle \semi \rangle / H)$.
In the self-heating scenario, however, the DM freeze-out is delayed, because the DM temperature shortly increases relative to $T_{\phi}$ and enhances the backward semi-annihilation process.
This can be seen from Fig.~\ref{fig:delay_fo}, where we present numerically computed $\Delta = Y_{\chi} - {\cal J}Y_{\chi}^{\rm eq}$ as a function of $m_{\chi} / T_{\phi}$.
This delay results only in an ${\cal O}(10)$\% change of the final DM relic abundance.

It may be nontrivial why semi-annihilation still affects the evolution of $T_{\chi}$ even after the kinetic decoupling.
To illustrate temperature evolution more clearly, it is useful to consider the thermodynamics of the SM bath and DM bath. 
Here we assume $T_{\phi} = T_{\rm SM}$.
From the first law of thermodynamics, we find
\begin{align}
d \left( \rho_{\rm SM} V \right) &= d Q_{\rm SM} - p_{\rm SM} \, dV \,,
\\
d \left( \rho_{\chi} V \right) &= d Q_{\chi} - p_{\chi} \, dV + m_{\chi} \, dN_{\chi} \,.
\end{align}
Since we are interested in the dynamics after the freeze-out, we consider only forward semi-annihilation $\chi \chi \rightarrow \chi \phi$.
For a single forward semi-annihilation process,
\begin{align}
d Q_{\rm SM} &= - ( 2 - \gamma  ) m_{\chi} \, dN_{\chi},
\\
d Q_{\chi} &= - (\gamma -1 ) m_{\chi} \, dN_{\chi}\,,
\end{align}
where $\gamma = (5/4)[ 1 - m_{\phi}^{2} / (5m_{\chi}^{2}) ]$ is a Lorentz boost factor.
We see that the total energy of the whole system is conserved: $d Q_{\rm SM} + d Q_{\chi} + m_{\chi} d N_{\chi} = 0$.
To investigate how the DM temperature evolves, we note that $\rho_{\chi} = \left( m_{\chi} + 3 T_{\chi} / 2 \right) n_{\chi}$ and $p_{\chi} = T_{\chi} n_{\chi}$ in the non-relativistic limit of DM.
In addition, a small fraction of DM undergoes semi-annihilation even after the freeze-out, $d \ln N_{\chi} \simeq - (\Gamma_{\rm semi} / H) \, d\ln a$.
We find
\begin{align}
d \ln T_{\chi} \simeq \left[ - 2 
+  (\gamma - 1) \frac{2}{3} \frac{m_{\chi}}{T_{\chi}} \frac{\Gamma_{\rm semi}}{H} \right] d\ln a \,.
\label{aa}
\end{align}
The first term in the square brackets represents the adiabatic cooling due to the expansion of the Universe, while the second term is due to the energy injection through semi-annihilation.
If $T_{\chi} \propto 1/a^{2}$ as for free-streaming non-relativistic DM particles, then the second term increases with the expansion of the Universe as $\propto (m_{\chi} / T_{\chi} ) (\Gamma_{\rm semi} / H) \propto a$ and thus heats the DM particles. 
Hence, the DM temperature is determined by the balance between the adiabatic cooling and semi-annihilation heating as $T_{\chi} \propto 1/a$.

\section{Evolution equations of self-heating DM cosmological perturbations}
\label{sec:derivation_evolution_perturbations}
We derive the evolution equations of cosmological perturbations.
Taking the conformal Newtonian gauge [see Eq.~\eqref{eq:newtoniangauge}], we get the linearized Boltzmann equation given by
\begin{align}
\delta f' 
+ i \left( \frac{\vec{k} \cdot \vec{q}}{\epsilon} \right) \delta f
- \left[ 
\Phi' + i (\vec{k} \cdot \hat{q}) \frac{\epsilon}{q} \Psi 
\right] 
\frac{\partial \bar{f}}{\partial \ln q}
&= a \left( \frac{1}{E} C^{(1)} + \frac{1}{E} C^{(0)} \Psi \right) \,.
\end{align}
We expand $\delta f$ as a function of $\mu = \hat{k} \cdot \hat{q}$ in terms of the Legendre polynomial:
\begin{align}
\delta f (\tau, \, \vec{k}, \, q, \, \hat{q}) = 
\sum_{\ell = 0}^{\infty} (-i)^{\ell} (2 \ell + 1) F_{\ell} (\tau, \, k, \, q) P_{\ell} (\mu) \,.
\end{align}
Multiplying the Legendre polynomial by the linearized Boltzmann equation and integrating it with respect to $\mu$, we find the Boltzmann hierarchy as
\begin{align}
F'_{0} 
&=
- \frac{k q}{\epsilon} F_{1}
+ \Phi' \frac{\partial \bar{f}}{\partial \ln q} 
+ a \int_{-1}^{1} \frac{d \mu}{2} \,
\left( \frac{1}{E} C^{(1)} + \frac{1}{E} C^{(0)} \, \Psi \right)
\label{eq:F_0_eq_new} \,, \\
F'_{1} 
&=
- \frac{k q}{3 \epsilon} (2 F_{2} - F_{0})
- \frac{k \epsilon}{3 q} \Psi \frac{\partial \bar{f}}{\partial \ln q}
+ i a \int_{-1}^{1} \frac{d \mu}{2} \, P_{1} (\mu) \,
 \left( \frac{1}{E} C^{(1)} + \frac{1}{E} C^{(0)} \, \Psi \right)
\label{eq:F_1_eq_new} \,, \\
F'_{2} 
&=
- \frac{k q}{5 \epsilon} (3 F_{3} - 2 F_{1})
- a \int_{-1}^{1} \frac{d \mu}{2} \, P_{2} (\mu) \,
 \left( \frac{1}{E} C^{(1)} + \frac{1}{E} C^{(0)} \, \Psi \right)
\label{eq:F_2_eq_new} \,, \\ 
F'_{\ell} (\ell \geq 3)
&=
- \frac{k q}{(2 \ell + 1) \epsilon} [ (\ell + 1) F_{\ell + 1} - \ell F_{\ell - 1}]
+ (-i)^{-\ell} a \int_{-1}^{1} \frac{d \mu}{2} \, P_{\ell} (\mu) \,
 \left( \frac{1}{E} C^{(1)} + \frac{1}{E} C^{(0)} \, \Psi \right)
\label{eq:F_ell_eq_new} \,.
\end{align}

In terms of the DM fluid variables~\cite{Ma:1995ey}, one obtains
\begin{align}
\delta'_{\chi} =&
- 3 
\left( 
c_{s\chi}^{2} - \omega_{\chi} 
+ \frac{a}{3 {\cal H} \bar{\rho}_{\chi}} \int \frac{d^{3} \vec{p}}{(2 \pi)^{3}} E \frac{1}{E} C^{(0)}
\right) {\cal H} \delta_{\chi}
- 3 {\cal H} \pi_{\chi}
- (1 + \omega_{\chi}) \left( \theta_{\chi} + 3 \Phi' \right) \nonumber \\
& + \frac{a}{\bar{\rho}_{\chi}} 
\int \frac{d^{3} \vec{p}}{(2 \pi)^{3}} \, E \, \left(  \frac{1}{E} C^{(1)} +  \frac{1}{E} C^{(0)} \, \Psi \right)
\label{eq:delta_eq_new} \,, \\
\theta'_{\chi} =&
- (1 - 3 \omega_{\chi}) {\cal H} \theta_{\chi}
- \frac{\omega'_{\chi}}{1 + \omega_{\chi}} \theta_{\chi}
- k^{2} \sigma_{\chi}
+ k^{2} \Psi
+ \frac{1}{1 + \omega_{\chi}} k^{2} (c_{s\chi}^{2} \delta_{\chi} + \pi_{\chi}) \nonumber \\
& + \frac{i a k}{\bar{\rho}_{\chi}(1+\omega_{\chi})} 
\int \frac{d^{3} \vec{p}}{(2 \pi)^{3}} \, p  P_{1}(\mu) \, \left( \frac{1}{E} C^{(1)} + \frac{1}{E} C^{(0)} \, \Psi \right) 
\label{eq:theta_eq_new} \,, \\
\pi'_{\chi}=& 
- \left( 2 - 3 c_{s}^{2} - 3 \omega_{\chi} - \alpha_{0} \omega_{\chi} 
+ \frac{a}{3 {\cal H} \bar{\rho}_{\chi}} \int \frac{d^{3} \vec{p}}{(2 \pi)^{3}} \, E \, \frac{1}{E} C^{(0)} \right) {\cal H} \pi_{\chi} 
\nonumber\\
& - \left( \frac{c_{s\chi}^{2 \prime}}{{\cal H} c_{s\chi}^{2}} + 2 - 3 c_{s\chi}^{2} - \alpha_{0} \omega_{\chi} \right) c_{s\chi}^{2} {\cal H} \delta_{\chi} \nonumber \\
& + \left( c_{s\chi}^{2} - \frac{\alpha_{2} \omega_{\chi}}{3} \right) (1 + \omega_{\chi}) \theta
+ 3 \left[ c_{s\chi}^{2} - \frac{5 \omega_{\chi}}{3} + \omega_{\chi} \left( c_{s\chi}^{2} + \frac{\alpha_{3} \omega_{\chi}}{3} \right) \right] \Phi' \nonumber \\
& + \frac{a}{\bar{\rho}_{\chi}} \int \frac{d^{3} \vec{p}}{(2 \pi)^{3}} \, \left( \frac{p^{2}}{3 E} - c_{s}^{2} E \right) \, \left( \frac{1}{E} C^{(1)} + \frac{1}{E} C^{(0)} \, \Psi \right) 
\label{eq:pi_eq_new} \,, \\
\sigma'_{\chi} =&
- (2 - 3 \omega - \alpha_{1} \omega) {\cal H} \sigma_{\chi}
+ \frac{4}{15} \alpha_{2} \omega_{\chi} \, \theta_{\chi}
- \frac{2 k}{5 \bar{\rho}_{\chi} ( 1 + \omega_{\chi}) } 
	\int \frac{d^{3} \vec{p}}{(2 \pi)^{3}} \, \frac{p^{3}}{E^{2}} \, F_{3}
\nonumber\\
& - \frac{2 a}{3 \bar{\rho}_{\chi} (1 + \omega_{\chi})}
	\int \frac{d^{3} \vec{p}}{(2 \pi)^{3}} \, \frac{p^{2}}{E} P_{2} (\mu) \,
	\left( \frac{1}{E} C^{(1)} + \frac{1}{E} C^{(0)} \, \Psi \right)
\label{eq:sigma_eq_new} \,.
\end{align}
We have decomposed the pressure perturbation into the isentropic and entropic parts as $\delta P_{\chi} = \bar{\rho}_{\chi} (c_{s\chi}^{2} \delta_{\chi} + \pi_{\chi})$, where the adiabatic sound speed squared is given by
\begin{align}
c_{s\chi}^{2} = \frac{\bar{P}_{\chi}'}{\bar{\rho}_{\chi}'} 
= \omega_{\chi} + \bar{\rho}_{\chi} \frac{\omega'_{\chi}}{\bar{\rho}'_{\chi}} \,.
\label{eq:sound_speed}
\end{align}
The dimensionless constants are defined as
\begin{align}
\alpha_{0} =& 
\left( \int \frac{d^{3} \vec{p}}{(2 \pi)^{3}} \, \frac{p^{2}}{E} \frac{p^{2}}{E^{2}} \, F_{0} \right) \bigg/ \left( \omega \int \frac{d^{3} \vec{p}}{(2 \pi)^{3}} \, \frac{p^{2}}{E} \, F_{0} \right) 
\label{eq:alpha_0} \,, \\
\alpha_{1} =& 
\left( \int \frac{d^{3} \vec{p}}{(2 \pi)^{3}} \, \frac{p^{2}}{E} \frac{p^{2}}{E^{2}} \, F_{2} \right) \bigg/ \left( \omega \int \frac{d^{3} \vec{p}}{(2 \pi)^{3}} \, \frac{p^{2}}{E} \, F_{2} \right) 
\label{eq:alpha_1} \,, \\
\alpha_{2} =&
\left( \int \frac{d^{3} \vec{p}}{(2 \pi)^{3}} \, p \frac{p^{2}}{E^{2}} \, F_{1} \right) \bigg/ \left( \omega \int \frac{d^{3} \vec{p}}{(2 \pi)^{3}} \, p \, F_{1} \right)
\label{eq:alpha_2} \,.
\end{align}

We consider cosmological perturbations entering the horizon well after the freeze-out of the DM number density.
$\omega_{\chi} \simeq \left( T_{\chi} / m_{\chi} \right)$, $c_{s\chi}^{2} \simeq \left( T_{\chi} / m_{\chi} \right) \left[ 1 - d\ln T_{\chi} / \left( 3 \, d\ln a \right) \right]$, and $F_{\ell}$ ($\ell \geq 2$) are suppressed by the low DM velocity.
Furthermore, $F_{\ell}$ ($\ell \geq 2$) are erased by self-scattering of DM.
In this limit, we find Eqs.~\eqref{eq:delta_ev_syn} -- \eqref{eq:pi_ev_syn}.
We have substituted $\alpha_{2} = 5$, which can be derived as follows.
In general, $F_{\ell} (\tau, \, k, \, q)$ can be expanded by a complete system of functions of $q$.
We can expand $F_{\ell} (\tau, \, k, \, q) = \sum_{n} 1 / (2 \pi a^{2} m_{\chi} {T}_{\chi})^{2 / 3} y^{\ell / 2} L_{n}^{\ell + 1/2} (y) \, e^{-y} {\cal F}_{n \ell}$ with $y = q^{2} / (2 \pi a^{2} m_{\chi} {T}_{\chi})$ and $L_{n}^{\alpha}$ being the associated Legendre function, as in Refs.~\cite{Bertschinger:2006nq, Binder:2016pnr}.
An advantage of this complete system is that $n = 0$ gives the dominant contribution in the non-relativistic limit.
Since $L_{0}^{\alpha} = 1$, one can find $\alpha_{2} = 5$.
One may also calculate other $\alpha$'s in a similar manner, while they do not appear in Eqs.~\eqref{eq:delta_ev_syn}--\eqref{eq:pi_ev_syn} and thus are not given here.
One solves Eqs.~\eqref{eq:delta_ev_syn}--\eqref{eq:pi_ev_syn} simultaneously with
\bea
\omega_{\chi}' &=& 
-2 {\cal H} \omega_{\chi} 
+ \frac{2}{3} (\gamma - 1) a \Gamma_{\rm semi} \,,
\label{eq:omega_eq}
\\
c_{s \chi}^{2}  &=& \frac{5}{3} \omega_{\chi}
-\frac{2}{9} (\gamma -1) \frac{a \Gamma_{\rm semi}}{\cal H} \,.
\label{eq:cs_eq}
\eea
which can be directly obtained from Eq.~\eqref{eq:temp}.
These relations are valid as long as $\Gamma_{\rm self}  / H \gtrsim 1$. 

By taking the synchronous gauge,
\begin{align}
ds^2 = a^2(\tau) \left[ - d\tau^2 + ( \delta_{ij}  + h_{ij} ) dx^i dx^j \right]\,,
\end{align}
one can derive the evolution equations given as
\begin{align}
\delta_{\chi}'
=&
- \theta_{\chi} -  \frac{h'}{2} 
\label{eq:delta_eq_nr_syn} \,, \\
\theta_{\chi}' 
=&
- {\cal H} \theta_{\chi}
+ k^{2} (c_{s \chi}^{2} \delta_{\chi} + \pi_{\chi}) 
\label{eq:theta_eq_nr_syn} \,, \\
\pi_{\chi}' 
=&
- 2 {\cal H} \pi_{\chi}
+ \left( c_{s \chi}^{2} - \frac{5}{3} \omega_{\chi} \right) \left[ - {\cal H} \left( 1 - \frac{{\cal H}'}{{\cal H}^{2}} \right) \delta_{\chi}
+ \theta_{\chi} + \frac{h'}{2} \right]
\label{eq:pi_eq_nr_syn} \,,
\end{align}
where $h$ denotes the trace of $h_{ij}$.
They are equivalent to those in the conformal Newtonian gauge under the gauge transformation [see Eq.~(27) of Ref.~\cite{Ma:1995ey}].
To check the consistency, one needs to note that $c_{s}^{2} {\cal H} k^{2} \alpha = c_{s}^{2} {\cal H} (h' + 6 \eta') / 2$ is negligible when compared to $\theta_{\chi}'$ for non-relativistic DM.

\begin{figure}
\centering
\begin{minipage}{0.45\linewidth}
\includegraphics[width=1.0\linewidth]{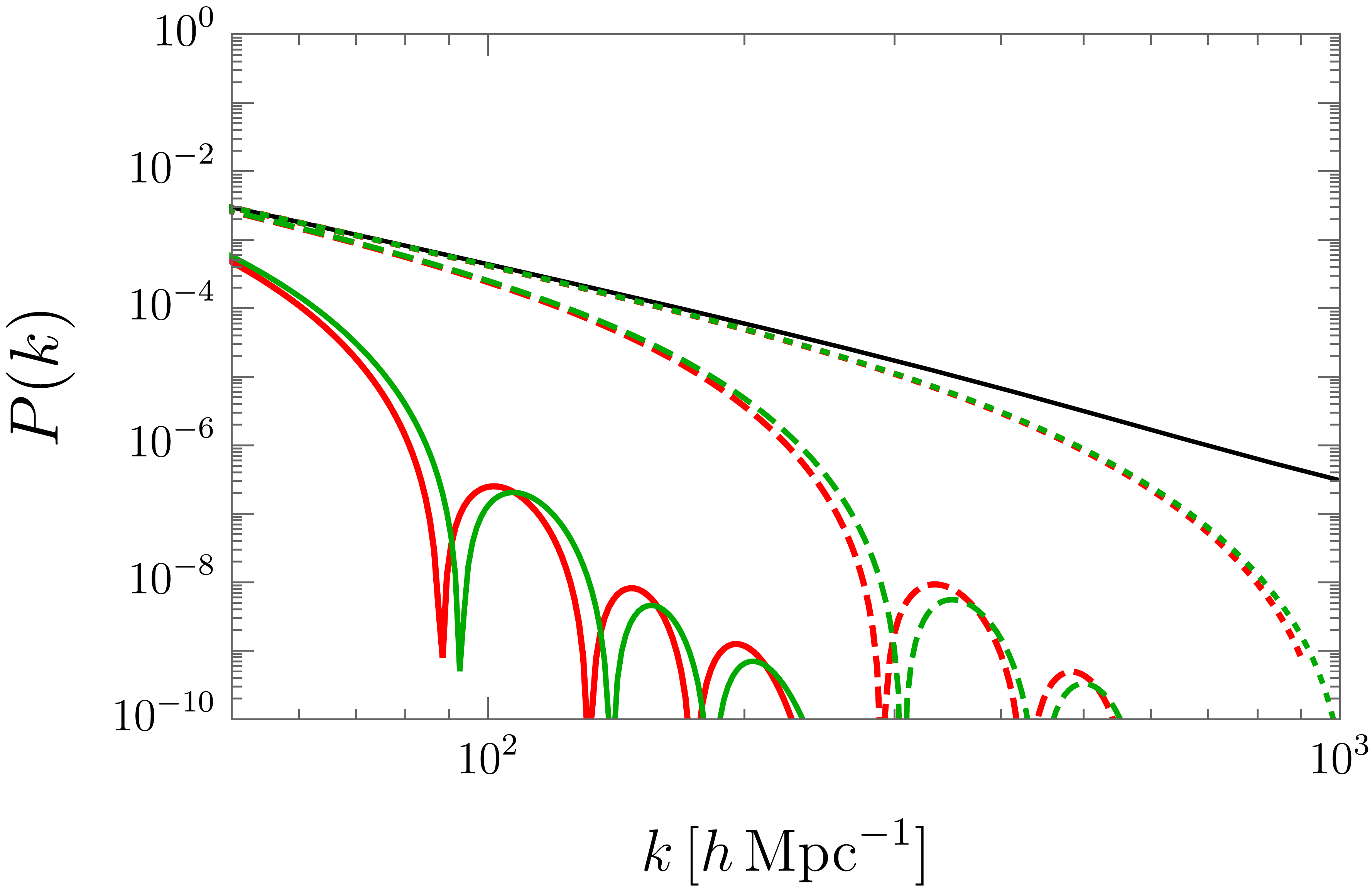}
\end{minipage}
\begin{minipage}{0.5\linewidth}
\includegraphics[width=1.0\linewidth]{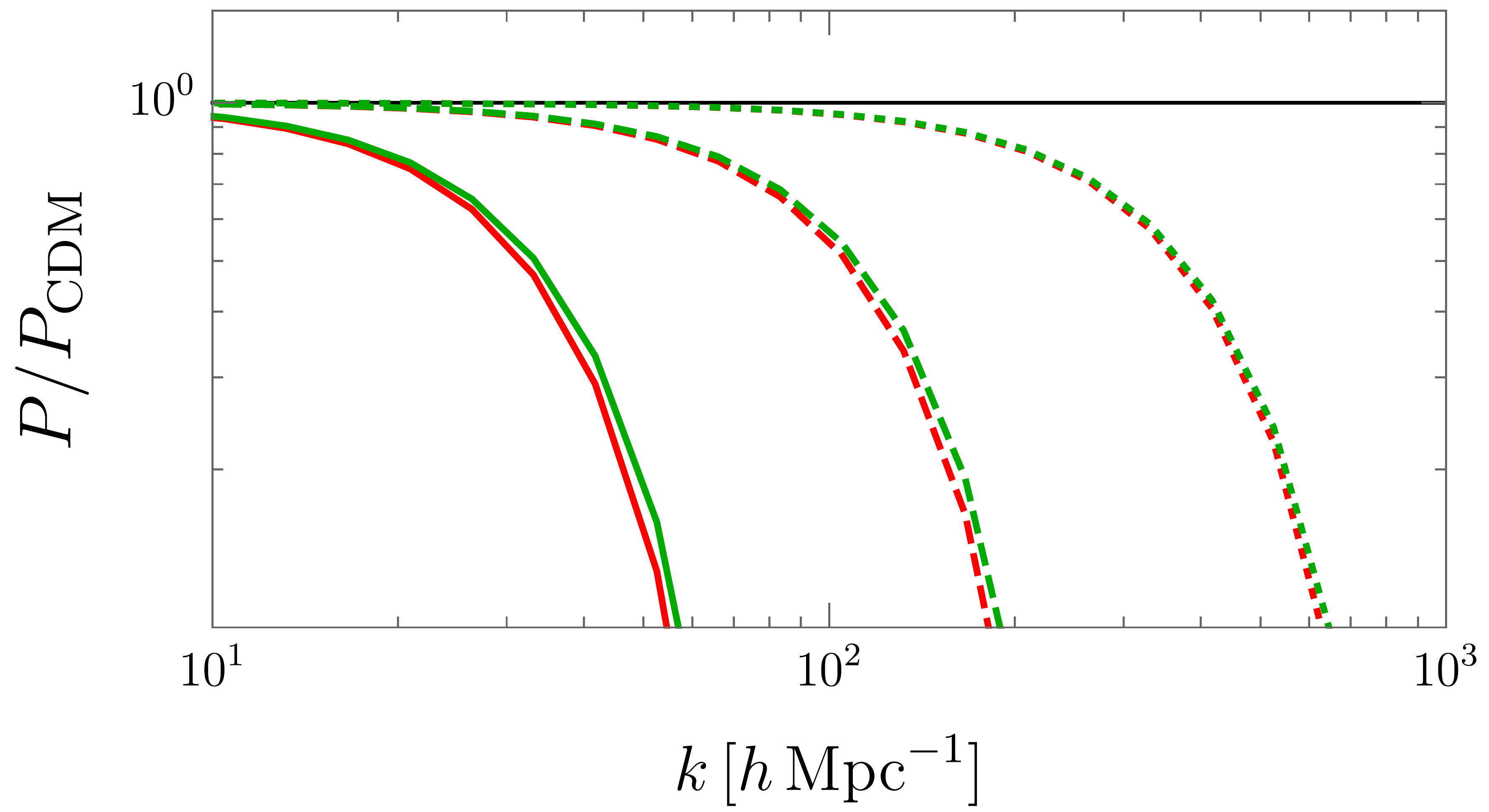}
\end{minipage}
\caption{%
(Left) Linear matter matter power spectrum when the decoupling of self-scattering is described by $(1- e^{-\Gamma_{\rm self} /H})$ ({\bf red}) and when it is described by $e^{-a/a_{\rm self}}$ ({\bf green)}.
(Right) Power spectrum in self-heating DM scenario relative to CDM. }
\label{fig:PS_dec}
\end{figure}

After decoupling of self-scattering [see Eq.\eqref{eq:selffreezeout}], energetic DM particles through semi-annihilation freely stream, while the majority of DM follows the Maxwell-Boltzmann distribution with the temperature $T_{\chi} \propto 1 / a^{2}$.
After that, Eqs.~\eqref{eq:delta_ev_syn}--\eqref{eq:pi_ev_syn} still describes the evolution of the majority of DM, although Eqs.~\eqref{eq:omega_eq}--\eqref{eq:cs_eq} are no longer valid.
For the decoupling of self-scattering, we take phenomenological approach, i.e., multiplying $(1-e^{-\Gamma_{\rm self}/H})$ to the equations of $\omega_{\chi}$ and $c_{s\chi}^{2}$ [see Eqs.~\eqref{eq:omega_eq_cut}--\eqref{eq:cs_eq_cut}]. 
We may introduce an alternative cutoff, for instance, like
\bea
\omega_{\chi}' &=& 
-2 {\cal H} \omega_{\chi} 
+ \frac{2}{3} (\gamma - 1) a \Gamma_{\rm semi}
\exp \left[ - \left( \frac{a}{a_{\rm self}} \right) \right] \,,
\\
c_{s \chi}^{2}  &=& \frac{5}{3} \omega_{\chi}
-\frac{2}{9} (\gamma -1) \frac{a \Gamma_{\rm semi}}{\cal H} 
\exp \left[ - \left( \frac{a}{a_{\rm self}} \right) \right] \,,
\eea
where $a_{\rm self}$ is the scale factor when $\Gamma_{\rm self} = H$.
In this case, the decoupling takes place more rapidly compared to the case presented in the main text, and the cutoff in the matter power spectrum appears for larger $k$.
See Fig.~\ref{fig:PS_dec} for the numerical difference in the two descriptions.
 
\bibliographystyle{utphys}
\bibliography{self-heating}

\end{document}